\documentclass{article}

\usepackage[british]{babel}  
\usepackage[scaled=0.86]{berasans}  
\usepackage[colorlinks=true, allcolors=blue, urlcolor=blue]{hyperref}  
\usepackage{graphicx} 
\usepackage[babel]{microtype}  
\usepackage{amsmath,amssymb,amsthm,bm,amsfonts,mathrsfs,bbm} 

\usepackage{xspace}  
\usepackage{pgfplots}
\usepackage{xcolor,colortbl}
\usepackage{tabularray}
\usepackage{hhline}
\usepackage{dsfont}
\usepackage{multirow}
\usepackage{pifont}
\usepackage{fancyhdr}
\usepackage{authblk}

\usepackage{array}
\usepackage{bigstrut}
\newcommand{\ket}[1]{| #1 \rangle}

\newcommand{\bea}{\begin{eqnarray}}
\newcommand{\eea}{\end{eqnarray}}

\newcommand{\xmark}{\ding{55}}

\begin{document}

\title{Contextual advantages across two-state discrimination strategies}


\author[1]{Kieran Flatt \thanks{kflatt@kaist.ac.kr}}
\author[1]{Joonwoo Bae \thanks{joonwoo.bae@kaist.ac.kr}}
\affil[1]{School of Electrical Engineering, Korea Advanced Institute of Science and Technology (KAIST), $291$ Daehak-ro, Yuseong-gu, Daejeon $34141$, Republic of Korea}

\maketitle

\begin{abstract}
Quantum state discrimination, alongside its other applications, has recently found use as a tool for witnessing generalised contextuality. In this article, we derive noncontextuality inequalities for both conclusive and inconclusive outcomes across various guessing strategies. For minimum-error discrimination, the advantage is in terms of the confidences of individual outcomes, while for unambiguous state discrimination, it is in terms of the average guessing probability. For maximum-confidence discrimination, we show that contextual advantages occur not only for the confidence but also their average, the guessing probability, as well as the inconclusive outcome rate. Our results show that contextual advantages are observed across all two-state discrimination schemes and figures of merit. We envisage that various quantum information applications based on state discrimination may offer advantages over noncontextual theories. \end{abstract}

\section{Introduction}

If gaps exist between the behaviour of classical and quantum systems, it is possible to take advantage of the differences for enhanced information processing. Contextuality, which tells us that measurements do not merely reveal pre-existing features of quantum states, has long been considered one such distinguishing feature. Despite this, it proved difficult for a long to know how to use this property for practical applications. This was in part due to the fact that non-projective measurements, which are required in realistic, noisy scenarios, were difficult to describe in the original framework of Kochen-Spekker contextuality \cite{Budroni2022KScontextuality}. For this reason, amongst others, a generalised form of contextuality was later put forward \cite{Spekkens2005Contextuality}. 
\par
Taking contextuality as a defining feature of quantum theory implies a natural definition of classicality as the set of statistics which can be generated by theories lacking this property. While Kochen-Spekken contextuality is reliant on the Hilbert space description of quantum theory, the generalised form is defined on the basis of observable probabilities. This allows one to define a set of noncontextual statistics, the violation of which signals a quantum advantage. With this in mind, recent work in the field of generalised contextuality has provided a series of noncontextual inequalities \cite{ Schmid2018NoncontextualInequalities, Lostaglio2020ContextualCloning}. These consist of proofs that the upper bound of a given operationally meaningful quantity is lower in theories lacking contextuality compared with the equivalent quantum figure of merit. Of particular importance here, as a fundamental communication primitive, is state discrimination \cite{Bae2015SD, Bergou2010SD, Barnett2009SD}.  It was first shown that quantum advantages may be witnessed in the setting of minimum-error state discrimination (MESD) \cite{Schmid2018ContextualSD} and subsequent works generalised this to aspects of unambiguous state discrimination (USD) and maximum confidence measurements (MCMs) \cite{Leifer2020, Shin2021, Flatt2022Contextual, RochiCarceller2022Randomness, Giordani2023, Wagner2023, Carceller2024}. Our aim in this paper is to complete this picture for two-state ensembles. We show that the ability to witness quantum advantages is universal across all forms of two-state discrimination, and in all measures of success. Note that, while ensembles with three-or-more states may be studied in this framework, it is not possible to construct \emph{all} such ensembles in a noncontextual theory \cite{Spekkens2005Contextuality}. Two-state ensembles are thus the only cases which can be studied in a general manner.
\par
Demonstrations of quantum advantage take one of two approaches. On the one hand, one can propose a measure of the average performance of a given task and then show that quantum implementations can attain a higher value than classical ones. The relevant notion of classicality in this case will be statistics that can be realised in non-contextual theories. In contrast, one can consider deterministic phenomena that can be witnessed in a single-shot scenario. The proof of the PBR theorem \cite{pusey2012reality}, for example, follows an analogous procedure by showing the existence of a task which can be performed perfectly in $\psi$-epistemic models but not in quantum theory \cite{Leifer2014Psi}.
\par
State discrimination, the problem of determining by measurement which state, taken from some ensemble, was prepared, allows one to witness both forms of advantage. MESD is concerned with the success probability observed on average over a run of experiments. USD and MCM, on the other hand, are concerned with inferring preparations from a single detection event, as what one learns does not depend on averaged data.  More generally, we may also ask about the single-shot properties of the former scheme as well as the averaged properties of the latter two.
\par
In order to do this, we note that the three different schemes may each be analysed according to three distinct figures of merit: the average probability of a correct detection event, the probability of conclusive measurement events, and the retrodictive confidence that a state was prepared given a measurement outcome. Optimisation of these three quantities gives arise to MESD, USD and MCM respectively. Less often noted is that the remaining figures of merit also play a role in each scenario. In this work, we show that they are also witnesses of contextuality. We show that, for MESD, the confidences of each individual outcome give rise to an associated contextual advantage. The same is true for the average guessing probability of USD. We also investigate the average guessing probability and inconclusive outcome rate of an MCM and derive further noncontextual inequalities.
\par
We begin by summarising both what is known so far and what the precise contributions are of this article. After reminding the reader of relevant background information, we then proceed with our main results: demonstrations that the confidences in MESD as well as the average success rates in USD and MCMs give rise to contextual inequalities. We close with remarks on the relevancy of these results to realistic experiments and applications.

\section{Summary of results}
State discrimination tasks, reviewed in more detail in the next section, are characterised by three basic figures of merit: the average guessing probability ${\rm P}_g$, the inconclusive outcome rate ${\rm P}_0$ and the set confidences $ {\rm C}(i)$ associated with measurement outcomes. These quantities can be associated with each of the three main versions of state discrimination: minimum error, unambiguous and maximum confidence. One also considers the resulting values of these quantities given optimisation of another.
\par
\begin{table*}
\centering
\begin{tblr}{width=0.8\textwidth,
cells={valign=m,halign=c}, 
colspec={| p{1.2cm} |  p{1.8cm} |  p{1.8cm} | p{1.8cm} | p{1.8cm} | p{1.8cm} | p{1.8cm}|},
}
\hline
 \SetCell[c=1,r=2]{} \textbf{Scheme} & \SetCell[c=3]{}  \textbf{Overview} & & & \SetCell[c=3]{} \textbf{Contextual advantages} \\
 \hline
  & Optimisation & Inc. Outcome? & All ensembles? & $P_g$ & $P_0$ & $C(i)$ \\
 \hline[1pt]
 \textbf{MESD} & $\max P_g$ & \xmark & \checkmark & Ref. \cite{Schmid2018ContextualSD} & 0 & This work \\
 \hline
 \textbf{USD} & $\min P_0$ & \checkmark & \xmark & This work & Ref. \cite{Flatt2022Contextual} & 1 \\
 \hline
 \textbf{MCM} & $\max {\rm C}(i) \, \forall \, i$ & \checkmark & \checkmark & This work & This work & Ref. \cite{Flatt2022Contextual} \\ 
 \hline
\end{tblr}
\caption{Three schemes may be used to implement quantum state discrimination. In this table, we display firstly the basic properties defining them. In the last three columns, we summarise the contextual advantages associated with each scheme. Cells containing citations were previously shown; cells containing numbers are definitional; cells stating `This work' are shown in the current work.}
\label{table:results}
\end{table*}

Table \ref{table:results} displays the various figures of merit which may be associated with each scheme. An advantage for the minimum error guessing probability was shown in Ref. \cite{Schmid2018ContextualSD}, and advantages for the maximum confidences and the inconclusive outcome rate were shown in Ref. \cite{Flatt2022Contextual}. We show, in the current article, the results for the remaining cases.

\section{Background} \label{sec:bg}
\subsection{State discrimination}

\begin{figure}[t!]
    \centering
    \hspace*{-0.5cm}                                                    
    \includegraphics[width=0.7 \linewidth]{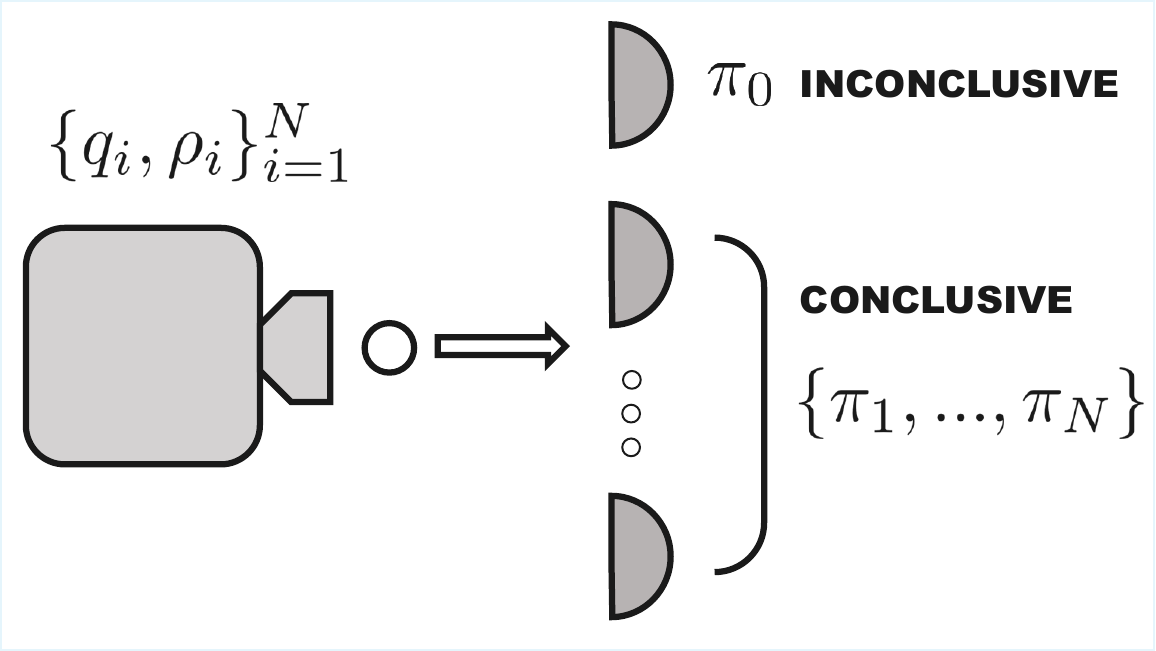}
    \caption{A general state discrimination scheme requires a part to send one state from an ensemble $\{q_i, \rho_i\}_{i=1}^N$ to another. The latter measures with a POVM which contains both conclusive ($\pi_1,...,\pi_N$) and inconclusive ($\pi_0$) outcomes.}
    \label{fig:comm}
\end{figure}

Any state discrimination task \cite{Bae2015SD, Barnett2009SD, Bergou2010SD} is characterised by a finite, discrete ensemble $\{ q_i, \rho_i \}_{i=1}^N$ of $N$ quantum states $\rho_i$ each prepared with a probability $q_i$. The average state produced is denoted $\rho=\sum_i^N q_i \rho_i $. One party, Alice, prepares a system in one of the states and then passes it on to a receiver, Bob. Bob's task is to determine the state using measurements, represented by a positive-operator-valued-measure (POVM) $\{ \pi_j \}$, such that  $0\leq \pi_i \leq 1$ and $\sum_j \pi_j=\mathbf{1}$, in which the outcomes are are associated with either a given state or an inconclusive detection event. This set-up is displayed in Figure \ref{fig:comm}. Bob will design his measurement to be optimal according to a chosen figure of merit, giving rise to MESD, USD and MCM depending on the choice. We summarise the key points of each scheme in Table \ref{table:results}.
\par
Three basic quantities can be used to measure the success of a protocol. One is the \emph{average guessing probability}, given by 
\begin{equation} \label{eq:guessprob}
    {\rm P}_g = \sum_{i=1}^N q_i {\rm P}_{M|P}(i|i) = \sum_{i=1}^N q_i  {\rm Tr}[\rho_i \pi_i ],
\end{equation}
which is the weighted average that measurement outcome $i$ occurs given that state $\rho_i$ was prepared. We employ (as throughout this article) the notation ${\rm P}_{M|P}(i|k)$ for the probability  that measurement $M$ resulted in outcome $i$ given that the preparation $P$ resulted in state $k$.
\par
The second figure of merit which is involved is the \emph{inconclusive outcome rate}. In certain protocols, it is necessary to include an $0$th outcome signifying that the measurement did not reveal anything about the prepare state. This may be by design or due to experimental flaws. The rate at which this happens is given by  
\begin{equation} \label{eq:incrate}
    {\rm P}_0 = {\rm Tr}[ \rho \pi_0].
\end{equation}
\par
Finally, the set of \emph{confidences}, 
\begin{equation} \label{eq:conf}
    {\rm C}(i) = {\rm P}_{P|M}(i|i) = \frac{q_i {\rm P}_{P|M}(i|i) }{ {\rm P}_M(i)} .
\end{equation}
are the retrodictive probabilities that a particular state was prepared given that the relevant measurement outcome was observed.
\par

Optimisation over each of these three quantities, with the potential of additional constraints, can be associated with a different form of state discrimination. If the average guessing probability, Eq. \ref{eq:guessprob} is maximised over an $N$ outcome measurement, one is said to perform minimum error state discrimination. In the simplest case of two states the maximum value is known to be the Helstrom bound \cite{Helstrom1976Quantum}.
\par
Unambiguous state discrimination \cite{Ivanovic1987USD, Dieks1988USD, Peres1998USD} refers to scenarios in which all conclusive outcomes, those labelled $1$ to $N$, determine the prepared state with certainty. This result is achieved by the $i$th POVM element projecting onto a subspace outside of the support of the span of the other states. A necessary condition, referred to as Chefles' criterion \cite{Chefles1998USD}, for such a measurement to be possible is that the states in the ensemble are all linearly independent. POVMs implementing this will typically include an inconclusive outcome, and minimising its occurence rate, Eq. \ref{eq:incrate}, will be optimal.
\par
A maximum confidence measurement \cite{Croke2006MCM, Mosley2006MCMExp, Lee2022MCM} is defined as one in which every confidence, Eq. \ref{eq:conf}, reaches its maximum value for all $i$. This is known to happen if each POVM element $\pi_i$ is a rank-one measurement proportional to the projector onto the eigenvector of $\rho^{-1} \rho_i \rho^{-1}$ with the highest associated eigenvalue. In general, one will maximise the set of confidences, and then minimise the inconclusive outcome rate associated with the optimal measurement. For scenarios involving two-dimensional systems, a semi-definite programming approach has been developed for this optimisation task \cite{Lee2022MCM}. Note that, under this approach, we can redefine USD as the subset of MCMs in which it is possible to attain ${\rm C}(i)=1$ for all $i$ in the ensemble.

\subsection{Generalised contextuality}
Different definitions of contextuality have been proposed and here we focus on the `generalised contextuality' approach developed by Spekkens \cite{Spekkens2005Contextuality}. This framework shows that operational theories subject to a constraint of non-contextuality cannot reproduce all of statistics given by quantum theory. 
\par
An operational theory represents experimental procedures as mathematical functions on an ontic state space $\Lambda$, in which each state is associated with a point $\lambda$ encoding all physical properties of the system. An experiment is then decomposed into preparations, transformations and measurements. Preparations $P$ are defined as epistemic states, $\mu_{j}(\lambda)$, probability distributions over $\Lambda$. Measurements $M$ are treated as sets of positive response functions $\{ \xi_{i|M} (\lambda) \}$ normalised so that $ \sum_i \xi_{i|M} (\lambda) = 1 \, \, \forall \lambda $. The probability of a measurement outcome $i$ given measurement $M$ and preparation $j$ in such models is given by
\begin{equation}
   {\rm P}_{M|P}(i | j) = \int_{\Lambda} d\lambda \mu_j (\lambda) \xi_{i|M} (\lambda) .
\end{equation}
Two epistemic states are said to be \emph{preparation equivalent} if every choice of measurement gives the same statistics for both states:
\begin{equation}
    \int_{\Lambda} d\lambda \mu_{1} (\lambda) \xi_{i|M} (\lambda) = \int_{\Lambda} d\lambda \mu_{2} (\lambda) \xi_{i|M} (\lambda) \, \, \, \forall \{i|M\} .
\end{equation}
We can now define a noncontextual theory: a theory is said to be preparation noncontextual if it assigns the same epistemic state to all equivalent preparations in a model. A similar definition holds for measurement noncontextuality.
\par
The final piece of terminology required is a measure for the distance between two pure states. In an operational model this is called the confusability, $c_{1,2}$, 
\begin{equation}
    \begin{split}
        c_{1,2} &= \int_{\Lambda} d\lambda \mu_2 (\lambda) \xi_1 (\lambda) \\
        &= \int_{ {\rm supp}[ \mu_1]} d \lambda \mu_2(\lambda)
    \end{split}
\end{equation}
where the second line follows from imposing noncontextuality and $\xi_1(\lambda)$ is a response function giving ${\rm P}_{M|P}(1|1)=1$. This function is symmetric such that $c_{1,2} = c_{2,1} $. When comparing noncontextual theories with quantum theory, this value is to be interpreted as the squared overlap of the state. That is, if the quantum state $\ket{\psi_1}$ and $\ket{\psi_2}$ are associated with the epistemic states $\mu_1(\lambda)$ and $\mu_2(\lambda)$ respectively, then the two theories are compared by fixing $c_{1,2} = | \langle \psi_1 | \psi_2 \rangle |^2$.
\par
Two additional assumptions are typically made of the epistemic model. One is that, for each pure state there exists a deterministic response function. The second is that probabilistic mixing over quantum states maps onto probabilistic mixing over the associated epistemic states:
\begin{equation}
    p \rho_1 + (1-p) \rho_2 ~~\mapsto ~~p \mu_1 (\lambda) + (1-p) \mu_2 (\lambda),
\end{equation}
in which $0\leq p\leq 1$ is a probability. These assumptions are used to construct a mapping between quantum theory and any operational model, whether noncontextual or otherwise.  
\section{Contextual advantages in state discrimination} \label{sec:known}

We begin in earnest by reviewing the previously known contextual advantages in state discrimination.
\par
\subsection{MESD: Success Probability}

The first contextual advantage to be shown in a state discrimination task was the guessing probability of a minimum error measurement \cite{Schmid2018ContextualSD}. The specific scenario investigated was an equiprobable two-state discrimination task, represented in quantum theory by two pure states $ | \psi_1 \rangle $ and $|\psi_2 \rangle $. The quantum upper bound of this problem is given by the well-known Helstrom bound \cite{Helstrom1976Quantum}
\begin{equation}
    {\rm P}^{(Q)}_g = \frac{1}{2} \left( 1 + \sqrt{1 - | \langle \psi_1 | \psi_2 \rangle |^2 } \right).
\end{equation}
This is compared with the maximum probability of distinguishing two preparations $\mu_1(\lambda)$ and $\mu_2(\lambda)$ with a confusability $c_{1,2}$ in a noncontextual theory. 
\par
In order to impose the constraint of noncontextuality, preparation equivalence is required. There, however, do not exist any equivalences if we invoke only two pure states. In order to get around this, the existence of a `mirror ensemble' $ \{ |\overline{\psi}_1 \rangle, | \overline{\psi}_2 \rangle \}$ consisting of the pair of states orthogonal to the initial ensemble $ \langle \psi_i | \overline{\psi}_i \rangle = 0 $ is assumed. With this addition we now have an equivalence between alternative preparations of the maximally mixed state:
\begin{equation}
    \frac{1}{2} | \psi_1 \rangle \langle \psi_1 | +  \frac{1}{2} | \overline{\psi}_1 \rangle \langle \overline{\psi}_1 | =  \frac{1}{2} | \psi_2 \rangle \langle \psi_2 | +  \frac{1}{2} | \overline{\psi}_2 \rangle \langle \overline{\psi}_2 |
\end{equation}
which then implies that in a noncontextual model the following constraint will apply: 
\begin{equation}
    \frac{1}{2} \mu_1 (\lambda) + \frac{1}{2} \overline{\mu}_1 (\lambda) =  \frac{1}{2} \mu_2 (\lambda) + \frac{1}{2} \overline{\mu}_2 (\lambda) 
\end{equation}
in which $ \overline{\mu}_i (\lambda) \cong | \overline{\psi}_i \rangle \langle \overline{\psi}_i |$. This relation constrains the relationship between the epistemic states in the considered model. In particular, one finds that
\begin{equation}
    \mu_1 (\lambda) = \mu_2 (\lambda) \, \, \, \forall \, \, \lambda \in {\rm supp}[ \mu_1] \cap {\rm supp}[\mu_2] .
\end{equation}
Application of this constraint along with the assumption that all epistemic states are associated with a deterministic response function then allows one to bound the guessing probability. The result is that
\begin{equation}
    {\rm P}_g^{(NC)} = 1 - \frac{c_{1,2}}{2} 
\end{equation}
which is strictly smaller than the Helstrom bound. 

\subsection{USD: Success Probability and Inconconclusive Outcome Rate }
We now turn our attention to the quantum advantages associated with unambiguous state discrimination, which were first demonstrated in Ref. \cite{Flatt2022Contextual}. Full technical details can be found in that article. These quantum advantages are associated with the inconclusive outcome rate, which is strictly higher in a non-contextual theory, and the guessing probability. We begin by noting that, as the probability of incorrectly guessing will be equal to zero, these two quantities are directly related to one another in USD, $P_g = 1 - P_0$, and it suffices to consider just one property. 
\par
It is trivial to show that in the binary discrimination scenario, USD is possible in a noncontextual theory. The response functions which satisfy these will be of the form 
\begin{equation}
    \xi_{i|M_{USD}} (\lambda) = 
    \begin{cases}
        \gamma, & \lambda \in {\rm supp}[\overline{\mu}_{i+1}] \\
        0, & {\rm otherwise}
    \end{cases}
\end{equation}
where addition in the subscripts is modular and $0 < \gamma \leq 1$ and $M_{USD}$ labels the measurement. Such response functions are equivalent to POVM elements of the form $\pi_i = \gamma | \psi_{i+1} \rangle \langle \psi_{i+1}  | $. This observation is furthermore supported by a recent theorem \cite{Zhang2025} stating that a sufficient condition for a set of states to be preparable in a preparation-noncontextual model is that they are linearly independent. This is in fact the same criterion as that for states to be unambiguously discriminable.

\par
One can then optimise the choice of $\gamma$ in order to minimise the inconclusive outcome rate, given that the response functions are constrained by preparation noncontextuality. The `mirror ensemble' approach, as taken in the minimum error calculation, is again used and the result is found to be
\begin{equation}
    {\rm P}_0^{(NC)} = \frac{1}{2}(1 + c_{1,2}).
\end{equation}
This value is strictly larger than or equal to the inconclusive outcome rate for quantum USD. In this manner, the contextuality of state discrimination was extended to USD. 
\par
\subsection{MCM: Confidences }
The next result to review is the advantage associated with the confidence in MCMs. As noted in the background, MCMs and USD coincide if the latter form is possible. The simplest case in which it is not is a dephasing of the binary pure state ensemble. That is, two states with the form 
\begin{equation} \label{eq:noisyqstates}
    \rho_i = (1-p) |\psi_i \rangle \langle \psi_i | + p \frac{\mathds{1}}{2} \, \, \, i=1,2
\end{equation}
where $0 \leq p \leq 1$ quantifies the noise existing in each state. The quantum problem can be solved using typical methods and the maximum confidence for each state is found to be
\begin{equation} \label{eq:mcq}
    {\rm C}^{(Q)}(i) = \frac{1}{2} \left(1 + \frac{(1-p) \sqrt{1 - | \langle \psi_1 | \psi_2 \rangle |^2}}{\sqrt{1 - (1-p)^2 | \langle \psi_1 | \psi_2 \rangle |^2}} \right).
\end{equation}
Using the same methods as discussed above, the noncontextual bound to the confidence can also be calculated. This is found to be
\begin{equation} \label{eq:mcnc}
    {\rm C}^{(NC)}(i) = \frac{1}{2} \left( 1 + \frac{(1-p)(1-c_{1,2})}{1 - (1-p)c_{1,2}} \right)
\end{equation}
which again is smaller than the equivalent quantum confidence seen above.
\par
It was shown in these three results that quantum advantages are common to different forms of state discrimination. What we will see in the remaining sections is that this advantage can also be observed for other figures of merit. 

\section{Characterising contextual advantages} \label{sec:new}
In this section, we derive further noncontextual inequalities for each state discrimination strategy. We do this by performing the following set of optimisation tasks in both quantum and noncontextual theories. First, we maximise the guessing probability and then calculate the confidences of the outcomes. Second, we optimise the set of confidences and  find the minimum inconclusive outcome rate and guessing probability for the optimal measurement.

\subsection{MESD: Confidences}

Let us consider two pure states that are prepared with arbitrary {\it a priori} probabilities $q_1, q_2 >0$, that is, the ensemble $\{ q_i, |\psi_i \rangle \langle \psi_i | \}_{i=1}^2$. We construct the measurements that maximise the guessing probability and then calculate confidences of these measurements.

Let us begin with the quantum scenario. Two pure states may be parameterised without loss of generality as 
\begin{equation}
\begin{split}
    \ket{\psi_1} &= \cos(\frac{\theta}{2}) \ket{0}- \sin(\frac{\theta}{2}) \ket{1} \\
    \ket{\psi_2} &= \cos(\frac{\theta}{2}) \ket{0} + \sin(\frac{\theta}{2}) \ket{1} 
\end{split}
\end{equation}
for which the squared overlap is $|\langle \psi_1|\psi_2\rangle|^2 = \cos\theta  = c_{1,2}$. 
An optimal measurement that maximises the guessing probability can be obtained by finding the eigenvectors of the operator, $q_1 |\psi_1 \rangle \langle \psi_1 | - q_2 |\psi_2 \rangle \langle \psi_2 |$. The POVM which results in maximum success is then given by $\pi_i = |\phi_i\rangle\langle\phi_i|$, with
\begin{equation}
\begin{split}
    \ket{\phi_1} &= \cos(\frac{\phi}{2}) \ket{0}- \sin(\frac{\phi}{2}) \ket{1} \\
    \ket{\phi_2} &= \cos(\frac{\phi}{2}) \ket{0} + \sin(\frac{\phi}{2}) \ket{1} 
\end{split}
\end{equation}
and
\begin{equation} \label{eq:mesdphi}
\tan(\frac{\phi}{2}) = \frac{(q_2-q_1)\cos(\theta)-\sqrt{1-4q_1q_2 \cos^2(\theta)}}{\sin(\theta)}.
\end{equation}
The resulting maximum success probability is
\begin{equation}
P_g^{(Q)} = \frac{1}{2} \left( 1 + \sqrt{ 1 – 4 q_1 q_2 \cos^2 (\theta) }\right)
\end{equation}
The maximum confidence given by MESD in quantum theory can therefore be evaluated straightforwardly for
\begin{equation} \label{eq:mesdconfq}
\begin{split}
    {\rm C}^{(Q)}(1) &= \frac{q_1 \cos^2(\frac{\theta+\phi}{2})}{q_1\cos^2(\frac{\theta+\phi}{2}) + q_2 \cos^2(\frac{\theta-\phi}{2}) } \\
    C^{(Q)}(2) &= \frac{q_2 \cos^2(\frac{\theta+\phi}{2})}{q_1\cos^2(\frac{\theta-\phi}{2}) + q_2 \cos^2(\frac{\theta+\phi}{2}) },
    \end{split}
\end{equation}
in which $\phi$ is evaluated using Eq. \ref{eq:mesdphi}. Here, unlike for the equiprobable case, the confidences associated with each outcome differ. 

We compare this with the equivalent quantity in a noncontextual. We first calculate the upper-bound for the MESD success probability for the relevant ensemble. This was first done in Ref. \cite{Shin2021}, where
\begin{equation} \label{eq:ncmesd}
{\rm P}_g^{(NC)} = 1 – (1 - \max \{ q_1, q_2 \}) c_{1,2}. 
\end{equation}
This figure of merit is seen to display a contextual advantage. We require, alongside the upper-bound, the particular set of response functions that reach this bound in order to evaluate the confidence. This can be done constructively. The set of two-outcome measurements in a two-state discrimination task can be represented by two sets of response functions. First, $M_1$ is given by
   \begin{equation} 
    \begin{split}
    \xi_{1|M_1} (\lambda) &= 
    \begin{cases}
        1, & \lambda \in {\rm supp}[\mu_{1}] \\
        0, & {\rm otherwise}
    \end{cases} \\
    \xi_{2|M_1} (\lambda) &= 
    \begin{cases}
        0, & \lambda \in {\rm supp}[\mu_{1}] \\
        1, & {\rm otherwise}
    \end{cases}
    \end{split}
\end{equation}
and $M_2$ is given by
\begin{equation}
    \begin{split}
    \xi_{1|M_2} (\lambda) &= 
    \begin{cases}
        0, & \lambda \in {\rm supp}[\mu_{2}] \\
        1, & {\rm otherwise}
    \end{cases} \\
    \xi_{2|M_2} (\lambda) &= 
    \begin{cases}
        1, & \lambda \in {\rm supp}[\mu_{2}] \\
        0, & {\rm otherwise}
    \end{cases}
    \end{split}
\end{equation}
Any set of response functions is then given by a convex combination of these two measurements. Operationally speaking, this means that we can implement any two-outcome measurement in a non-contextual theory by implementing $M_1$ with probability $\omega$ and $M_2$ with probability $1-\omega$. The response function resulting from this procedure will be 
\begin{equation} \label{eq:NCmix}
    \begin{split}
    \xi_{1|M_\omega} (\lambda) &= 
    \begin{cases}
        \omega, & \lambda \in {\rm supp}[\mu_{1}] \cap {\rm supp}[\mu_{2}] \\
        1, & \lambda \in {\rm supp}[\mu_{1}] \cap {\rm supp}[\overline{\mu}_{2}] \\
        0, & \lambda \in {\rm supp}[\overline{\mu}_{1}] \cap {\rm supp}[\mu_{2}] \\
        1-\omega, & \lambda \in {\rm supp}[\overline{\mu}_{1}] \cap {\rm supp}[\overline{\mu}_{2}]
    \end{cases} \\
    \xi_{2|M_\omega} (\lambda) &= 
    \begin{cases}
        1-\omega, & \lambda \in {\rm supp}[\mu_{1}] \cap {\rm supp}[\mu_{2}] \\
        0, & \lambda \in {\rm supp}[\mu_{1}] \cap {\rm supp}[\overline{\mu}_{2}] \\
        1, & \lambda \in {\rm supp}[\overline{\mu}_{1}] \cap {\rm supp}[\mu_{2}] \\
        \omega, & \lambda \in {\rm supp}[\overline{\mu}_{1}] \cap {\rm supp}[\overline{\mu}_{2}]
    \end{cases}
    \end{split}
\end{equation}
and we now evaluate the success probability using this measurement, in order to optimise over $\omega$: 
\begin{equation}
    {\rm P}_g^{(NC)} = q_1 \int_{\Lambda} d\lambda \mu_1(\lambda) \xi_{1|\omega}(\lambda) + q_2 \int_{\Lambda} d\lambda \mu_2(\lambda) \xi_{2|\omega}(\lambda)  
\end{equation}
We can evaluate this expression using Eq. \ref{eq:NCmix} and begin from the first term:
\begin{equation}
    \begin{split} 
    \int_{\Lambda} d\lambda \xi_{1|M_\omega} (\lambda) \mu_1 (\lambda) = \omega & \int_{{\rm supp}[\mu_{1}] \cap {\rm supp}[\mu_{2}] } d \lambda \mu_1(\lambda) \\
    &+ \int_{{\rm supp}[\mu_{1}] \cap {\rm supp}[\overline{\mu}_{2}]} d\lambda \mu_1 (\lambda) ,
    \end{split}
\end{equation}
which can be simplified upon noticing the following relation:
\begin{equation}
\begin{split}
    \int_{{\rm supp}[\mu_{1}] \cap {\rm supp}[\mu_{2}] } d \lambda \mu_1(\lambda) = &\int_{{\rm supp}[\mu_{1}] \cap {\rm supp}[\mu_{2}] } d \lambda \mu_1(\lambda) \\
    &+ \int_{{\rm supp}[\overline{\mu}_{1}] \cap {\rm supp}[\mu_{2}] } d \lambda \mu_1(\lambda) \\
    &= \int_{ {\rm supp}[\mu_{2}] } d\lambda \mu_1 (\lambda) \\
    &= c_{1,2} .
    \end{split}
\end{equation}
and an analogous procedure gives the other values. The resulting success probability is 
\begin{equation}
P^{(NC)}_g = 1 - \left(1 - \omega q_1 - (1 - \omega) q_2 \right) c_{1,2} \leq 
1 - (1 - \max\{ q_1, q_2\}) c_{1,2} 
\end{equation}
which reproduces the upper bound in Eq. \ref{eq:ncmesd}. The measurement which maximises this is to choose $\omega=1$, that is, implement $M_1$, if $q_1>q_2$. If, on the other hand, $q_2 > q_1$, then the optimal choice is $\omega = 0$, that is, to implement $M_2$. The quantum analogy of these two measurements would be to project onto the basis defined by $|\psi_1\rangle$ ($|\psi_2\rangle$) for $M_1$ ($M_2$). Note that, if $q_1 = q_2$ then all choices of $\omega$ will give the same, previously known, upper bound for the non-contextual success probability. 
We are now in a position to evaluate the confidence for each measurement outcome. We may assume without loss of generality that we are in a scenario in which $q_1>q_2$, and so use $M_1$ to maximise the success probability. 

\begin{figure}[t]
    \centering  
    \includegraphics[width=0.8 \linewidth]{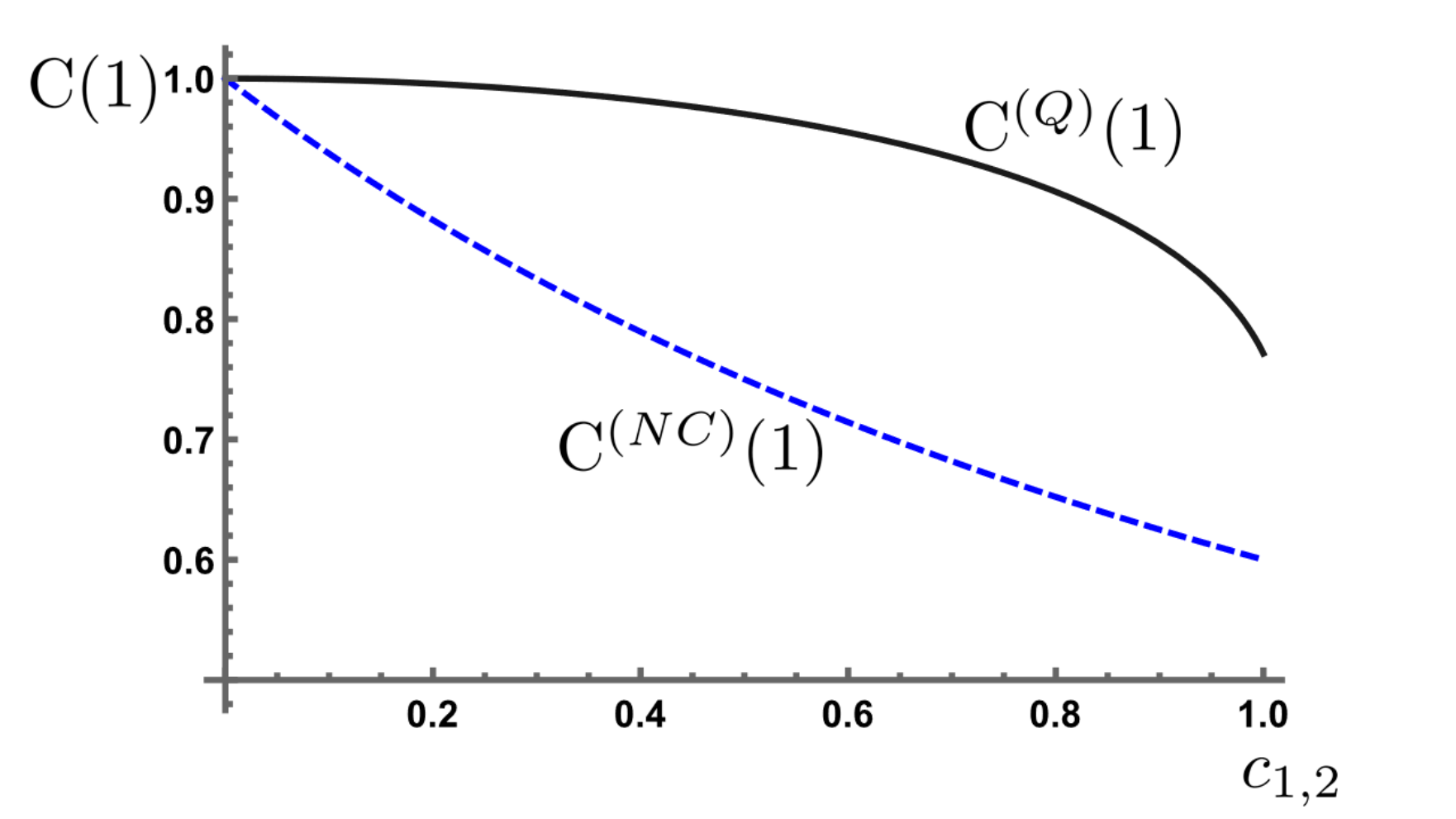}
    \caption{The confidence ${\rm C}(1)$ in both quantum (solid black line) and noncontextual (dashed blue line) theories for pure states of overlap $c_{1,2}$ where state 1 is prepared with {\it a priori} probability $q_1=0.6$.}
    \label{fig:conf1}
\end{figure}

The evaluation of the confidences using this measurement can be performed 
\begin{equation}
\begin{split}
C^{(NC)}(1) &= \frac{ q_1}{c_{1,2} + (1- c_{1,2}) q_1 } \\
C^{(NC)}(2) &= 1,
\end{split}
\end{equation}
which are distinct from the quantum counterparts in Eq. \ref{eq:mesdconfq}. We plot the quantum and noncontexual values of $C(1)$ in Fig. \ref{fig:conf1}. 
 
We observe that, out of two arms for outcomes $1$ and $2$, only the arm detecting a state with larger {\it a priori} probability may contain contextual advantage whereas the other not. This contrasts with the result in Ref. \cite{Flatt2022Contextual} where the {\it a priori} probablities are equal; the values of confidence for two outcomes are identical and thus both have a contextual advantage. One may point out that the state space structure may give rise to the distinction. The set of quantum states has extremal points continuously connected by state transformations, in contrast to which, that of a noncontextual theory has a finite number of extremal ones. Consequently, when {\it a priori} probabilities are not equal, optimal quantum measurements are also continuously transformed whereas their noncontextual counterparts are biased to some specific extreme points. 

\color{black}

\subsection{MCM: Inconclusive Outcome Rate}
We now turn our attention to the inconclusive rate in maximum confidence measurements, and will study the noisy discrimination scenario, when the aim in the quantum case is to discriminate the states in Eq. \ref{eq:noisyqstates} and the analogy of this in the ontological model is to discriminate the epistemic states
\begin{equation}
    \tilde{\mu}_i (\lambda) = (1-p) \mu_i (\lambda) + p \mu_{ \mathds{1}/2} (\lambda), \, \, \, \, \, i=1,2.
\end{equation}
Here, $\mu_i (\lambda) $ are two extremal epistemic states associated with the two pure states in the quantum framework and $\mu_{\mathds{1}/2} (\lambda)$ is the epistemic-state representation of the maximally mixed quantum state $\mathds{1}/2$. Note that in noncontextual theories the latter takes the same representation for all methods of preparation (i.e., all decompositions in the quantum framework). 
\par
The quantum task has been studied in detail \cite{Lee2022MCM}. If the states are parameterised as
\begin{equation}
\begin{split}
    \ket{\psi_1} &= \cos(\frac{\theta}{2}) \ket{0}- \sin(\frac{\theta}{2}) \ket{1} \\
    \ket{\psi_2} &= \cos(\frac{\theta}{2}) \ket{0} + \sin(\frac{\theta}{2}) \ket{1} 
\end{split}
\end{equation}
then optimal POVM elements will be rank-one and follow the structure $\pi_i = \alpha | \phi_i \rangle \langle \phi_i | \, \, i=1,2$, where $0\leq\alpha\leq1$ is a free parameter and
\begin{equation}
\begin{split}
    \ket{\phi_i} &= \cos(\frac{\phi}{2}) \ket{0} + (-1)^i \sin(\frac{\phi}{2}) \ket{1} \\
    \tan(\phi) &= p\cos(\theta)\sqrt{\frac{ 1-p^2 \cos^2 (\theta)}{1 - \cos^2(\theta)}} 
\end{split}
\end{equation}
\par
The important point here is to note that the optimal quantum measurement is a function of the amount of dephasing noise. From this construction we can write the inconclusive POVM element as $\mathds{1} - \pi_1 - \pi_2$ and minimise over the parameters $\alpha$ the inconclusive outcome rate, which is found to be
\begin{equation} \label{eq: quantump0}
{\rm P}^{(Q)}_0 = (1-p) | \langle \psi_1 | \psi_2 \rangle |.
\end{equation}
This is to be compared with the equivalent quantity in a noncontextual model of MCMs. The feature which we must use is that the noncontextually optimal measurement is represented by the same measurement as is used for unambiguous state discrimination:
\begin{equation}
\begin{split}
    \xi_{1|M_{USD}}(\lambda) &= 
    \begin{cases}
        \gamma_1, & \lambda \in {\rm supp}[\overline{\mu}_{2}] \\
        0, & {\rm otherwise}
    \end{cases} \\
     \xi_{2|M_{USD}} (\lambda) &= 
    \begin{cases}
        \gamma_2, & \lambda \in {\rm supp}[\overline{\mu}_{1}] \\
        0, & {\rm otherwise}
    \end{cases}
\end{split}
\end{equation}
where the response function representing an inconclusive outcome will be given by $\xi_{0|M_{USD}}(\lambda) = 1 - \xi_{1|M_{USD}} (\lambda) - \xi_{2|M_{USD}} (\lambda) $ and $0 < \gamma_i \leq 1$ for $i=1,2$. This response function in full will have the structure
\begin{equation} \label{eq:MCMincresp}
\begin{split}
    \xi_0 (\lambda) = 
    \begin{cases}
        1, & \lambda \in {\rm supp}[\mu_{1}] \cap {\rm supp}[\mu_{2}] \\
        1-\gamma_1, & \lambda \in {\rm supp}[\mu_{1}] \cap {\rm supp}[\overline{\mu}_{2}] \\
        1-\gamma_2, & \lambda \in {\rm supp}[\overline{\mu}_{1}] \cap {\rm supp}[\mu_{2}] \\
        1-\gamma_1 - \gamma_2, & \lambda \in {\rm supp}[\overline{\mu}_{1}] \cap {\rm supp}[\overline{\mu}_{2}]
    \end{cases} 
    \end{split}
\end{equation}

\begin{figure}[t]
    \centering  
    \includegraphics[width=0.8 \linewidth]{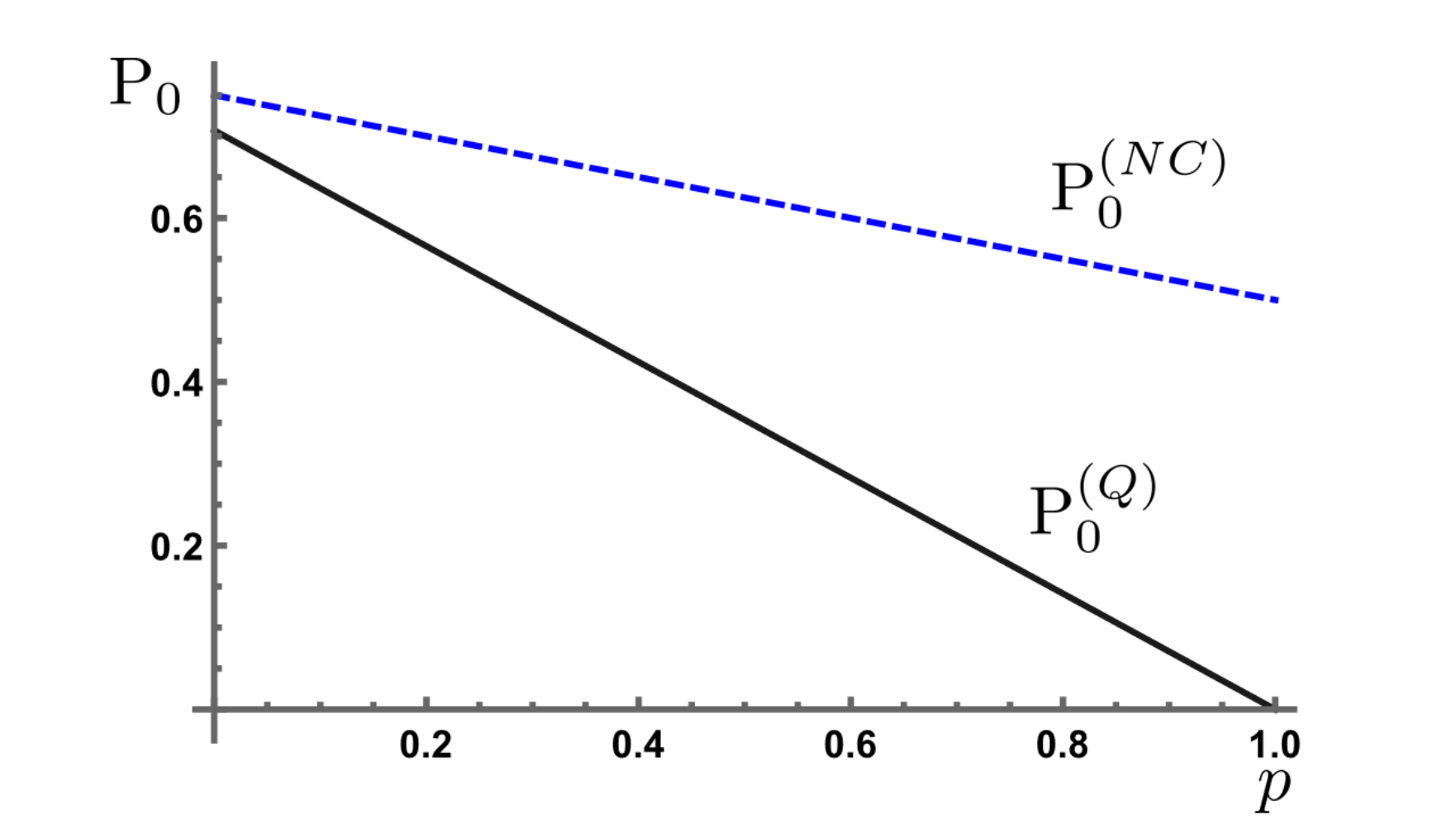}
     \includegraphics[width=0.8 \linewidth]{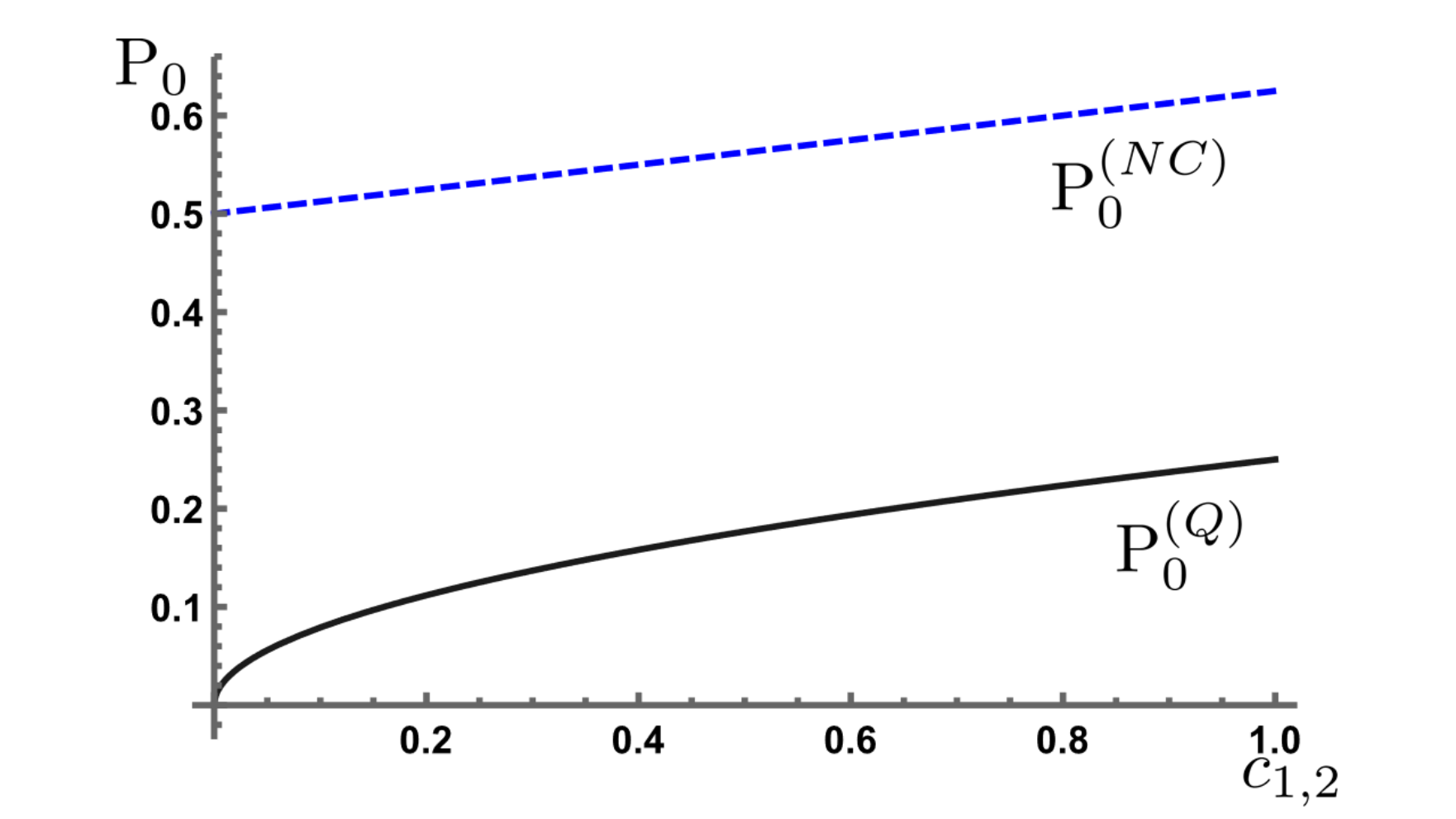}
    \caption{The inconclusive outcome probability $\mathrm{P}_0$ in both quantum (solid black line) and noncontextual (dashed blue line) theories for the task of discriminating two noisy states characterised by a noise parameter $p$ and overlap $c_{1,2}$ is plotted for varying $p$ (fixed overlap $c_{1,2}=1/2$) and varying $c_{1,2}$ (fixed noise $p=3/4$). }
    \label{fig:incrates}
\end{figure}

From this it is possible to calculate the inconclusive outcome rate. First note that the average density matrix for the noisy epistemic states will be
\begin{equation}
    \tilde{\mu}(\lambda) = \frac{1-p}{2} \mu_1 (\lambda) + \frac{1-p}{2} \mu_2 (\lambda) + p \mu_{\mathds{1}/2} (\lambda)
\end{equation}
Then the quantity we wish to evaluate is
\begin{equation}
\begin{split}
    {\rm P}^{(NC)}_0 &= \int_{\Lambda} d\lambda \tilde{\mu}(\lambda) \xi_{0|M_{USD}} (\lambda) \\
    &= \frac{1-p}{2} \int_{\Lambda} d\lambda \xi_{0|M_{USD}} (\lambda) \mu_1 (\lambda) + \frac{1-p}{2} \int_{\Lambda} d\lambda \xi_{0|M_{USD}} (\lambda) \mu_2 (\lambda) \\
    &+ p \int_{\Lambda}d\lambda \xi_{0|M_{USD}} (\lambda) \mu_{\mathds{1}/2} (\lambda)
\end{split}
\end{equation}
We can evaluate each of these integrals in turn using Eq. \ref{eq:MCMincresp}. First, we find
\begin{equation}
    \begin{split}
        \int_{\Lambda}d\lambda \xi_{0|M_{USD}} (\lambda) \mu_1 (\lambda)  
        &= \int_{{\rm supp}[\mu_1]\cap {\rm supp}[\mu_2]} d\lambda \mu_1 (\lambda) \\ 
        &+ (1-\gamma_1) \int_{ {\rm supp}[\mu_1]\cap {\rm supp}[\overline{\mu}_2]} d\lambda \mu_2(\lambda) \\
        &= c_{1,2} + (1-\gamma_1) c_{1, \overline{2}} \\
        &= 1 - \gamma_1 + \gamma_1 c_{1,2}
    \end{split}
\end{equation}
A similar process gives
\begin{equation} 
\int_{\Lambda} d\lambda \xi_{0|M_{USD}} (\lambda) \mu_2 (\lambda) = 1 - \gamma_2 + \gamma_2 c_{1,2}
\end{equation}
The final piece to evaluate is the integral over the maximally mixed epistemic state. In order to do this, we can decompose the epistemic state into either preparation, noting that the two will be the same due to the model's noncontextuality. We can thus evaluate:
\begin{equation}
    \begin{split}
        \int_{\Lambda}d\lambda \xi_{0|M_{USD}}(\lambda) \mu_{\mathds{1}/2} (\lambda) &= \int_{\Lambda} d\lambda \xi_{0|M_{USD}}(\lambda) \left( \frac{1}{2} \mu_1 (\lambda) + \frac{1}{2} \overline{\mu}_1 (\lambda) \right) \\
        &= 1 - \frac{1}{2} (\gamma_1 + \gamma_2) 
    \end{split}
\end{equation}
where the second line follows using similar techniques as above. 
\par
Bringing the previous few results together we now have
\begin{equation}
    {\rm P}^{(NC)}_0 = 1 - \frac{1}{2} \left( 1 - (1-p)c_{1,2} \right) \left( \gamma_1 + \gamma_2 \right)
\end{equation}
We need to find the minimum value of this subject to the constraints imposed on the non-negativity of the response function. Clearly, this function reaches a minimum when $\gamma_1 + \gamma_2$ as is allowed. Examining the inconclusive response function, Eq. \ref{eq:MCMincresp}, we can see that the piece $\xi_0(\lambda) $ on the domain $ \lambda \in {\rm supp}[\overline{\mu}_1] \cap {\rm supp}[\overline{\mu}_2] $  requires that $\gamma_1 + \gamma_2 \leq 1$ to be non-negative. Therefore we can take the upper bound $\gamma_1 + \gamma_2 = 1$ and the minimum inconclusive rate in a non-contextual model will be
\begin{equation}
    {\rm P}^{(NC)}_0 = \frac{1}{2} \left(1 + (1-p) c_{1,2} \right)
\end{equation}
Which is to be compared with Eq. \ref{eq: quantump0} using $c_{1,2} = |\langle \psi_1 | \psi_2 \rangle |^2$. We display the two functions in Fig. \ref{fig:incrates} with both fixed $c_{1,2}$ and fixed noise $p$ where it can be seen that the inconclusive rate for a noncontextual model is greater than or equal to that of quantum theory.

\subsection{MCM: Guessing probability}
The final piece which we fill in is the guessing probability, Eq. \ref{eq:guessprob}, of a maximum confidence measurement. We can recast the guessing probability into a form more amenable to the current task by writing it in terms of the confidences and outcome rates of the ensemble:
\begin{equation} \label{eq:pgmc}
    {\rm P}_{g} = \sum_i {\rm P}_{M}(i) {\rm C}(i),  
\end{equation}
which follows simply from the definition and Bayes rule. The index $i$ here runs over the states in the ensemble.  
\par
We can again note that for pure quantum states and extremal epistemic states, the MCM coincides with unambiguous state discrimination. For these cases, we need simply to repeat that the guessing probability in USD is $P_g = 1 - P_0$ and we have the same advantage shown above. For this reason, we need only to focus on the noisy ensemble discussed in the previous subsection. 
\par
\begin{figure}[t]
    \centering
    \hspace*{-0.5cm}                                                    
    \includegraphics[width=0.8 \linewidth]{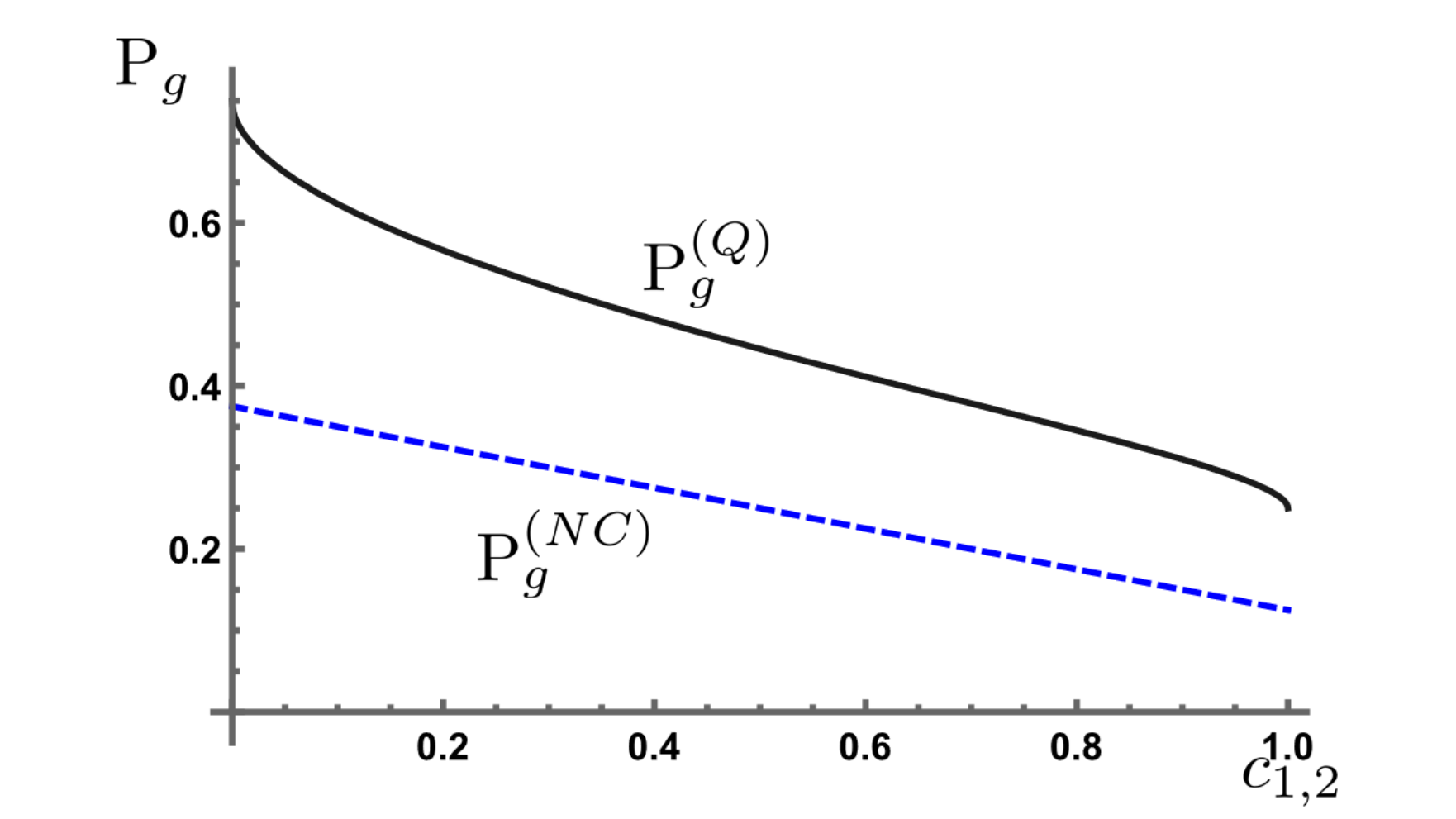}
    \caption{ The guessing probability by quantum (black solid line) and noncontextual (blue dashed line) implementations of a maximum confidence measurement on a binary, noisy ensemble characterised by probability $p=0.5$ as it varies with confusability $c_{1,2}$. }
    \label{fig:mcguessprob}
\end{figure}

The simplest way to demonstrate the contextual advantage here is to note that both confidences in a binary discrimination task are equal, as can be seen in Eqs. \ref{eq:mcq} and \ref{eq:mcnc}. We can therefore write ${\rm C}(1)={\rm C}(2)= {\rm C}$ and Eq. \ref{eq:pgmc} becomes
\begin{equation}
    \begin{split}
        {\rm P}_g &= \sum_{i} {\rm P}_M(i) {\rm C} \\
        &= (1 - {\rm P}_0 ) {\rm C}
    \end{split}
\end{equation}
Both $P_0$ and ${\rm C}$ have been previously calculated for both quantum and contextual theories, and both were seen to display a quantum advantage. As we have ${\rm P}_0^{Q} \leq {\rm P}^{(NC)}_0 $ and ${\rm C}^{(Q)}(i) \geq {\rm C}^{(NC)} (i) \, \, \forall i$, it trivially follows  that ${\rm P}^{(Q)}_g \geq {\rm P}^{(NC)}_g$ and a quantum advantage therefore exists also for this parameter. For completeness, the full expressions for these two quantities will be:
\par
\begin{equation}
    \begin{split}
        {\rm P}_g^{(NC)} &= \frac{1}{2} \left( 1 - \frac{p}{2} - (1-p)c_{1,2} \right) \\
        {\rm P}_g^{(Q)} &= \frac{1}{2} \Bigl( 1 - (1-p) | \langle \psi_1 | \psi_2 \rangle |  \\
       &   + (1-p) \sqrt{\frac{1 - (1-p)|\langle \psi_1 | \psi_2 \rangle |}{1 + (1-p) | \langle \psi_1 | \psi_2 \rangle | }} \sqrt{1 - | \langle \psi_1 | \psi_2 \rangle|^2} \Bigr)
    \end{split}
\end{equation}
\par
which are displayed in Fig. \ref{fig:mcguessprob}. 

\section{Discussion}
Demonstrating advantages of quantum systems over classical ones plays a central role in the developing of useful applications for quantum technologies. As state discrimination is the communication primitive underlying many applications, it is important to firmly state which property of quantum theory allow for it to perform above the classical limit. Here, we show that it is contextuality. While the contextual advantages associated with state discrimination were previously known, it was not clear which aspects were able to show a contextual advantage. The main lesson of this paper is that all figures of merit are reliant on this property. This was shown by demonstrating advantages for the guessing probability and inconclusive outcome rate in a maximum confidence measurement, as well as showing that under certain circumstances the confidence associated with a minimum error measurement also witness contextuality.
\par
Maximum confidence measurements, and the measurement confidences more generally, play an important role in this work. One reason is that, in experimental settings, one is forced to deal with preparation and measurement noise, including non-detection events. These can be naturally taken into account using the confidence as a figure of merit, unlike the average success probability. Another reason is that they allow us to witness two notions of quantum advantage. In a single-shot setting, we can make no claims about the average statistics. The confidence, however, is a well-defined quantity in this setting, and is able to see beyond-classical statistics.
\par
Our results should find use in experimental tests of contextuality, in which data are unavoidably noisy. It would be interesting to connect the inequalities found in state discrimination, a form of prepare-and-measure scenario, to nonlocal inequalities through the recent works bridging the two topics \cite{Plavala2024Contextuality, Wright2023NonlocalContextualtiy}, as well as understand the noise conditions under which contextuality can be witnessed \cite{Khoshbin2024NoncontextualityWitness} in state discrimination settings. As state discrimination plays an important role in many uses of quantum theory, for example randomness generation \cite{RochiCarceller2022Randomness, BohrBrask2017Randomness} and sequential communications \cite{Lee2024Sequential}, we, furthermore, believe that this work will find applications in practical use cases.

\section{Acknowledgments}
This work was supported by the National Research Foundation of Korea (RS-2024-00408613) and the Institute for Information \& Communication Technology Promotion (IITP) (RS-2023-00229524, RS-2025-02304540, RS-2025-25464876, RS-2025-25464616).


\end{document}